\documentclass[aps,pra,twocolumn,floatfix]{revtex4-1}
\usepackage[english]{babel}
\usepackage{bm,bbm,amsmath, amssymb,amsbsy, amsthm,graphicx,subfigure,color,txfonts,multirow}
\usepackage[colorlinks]{hyperref}
\hypersetup{
	bookmarksnumbered,
	pdfstartview={FitH},
	citecolor={blue},
	linkcolor={red},
	urlcolor={black},
	pdfpagemode={UseOutlines}
}

\begin{document}
\title{Towards Quantum Cybernetics}
 
\author{Davide Girolami$^{1}$, Rebecca Schmidt$^{2}$, and Gerardo Adesso$^{2}$} 
\affiliation{$^1\hbox{Clarendon Laboratory, Department of Physics, University of Oxford, Parks Road, Oxford OX1 3PU, United Kingdom}$}  
\affiliation{$^2\hbox{School of Mathematical Sciences, The University of Nottingham, University Park, NG7 2RD, Nottingham, United Kingdom}$ }

\begin{abstract}
 Cybernetics is a successful meta-theory to model the regulation of complex systems from an abstract information-theoretic viewpoint, regardless of the properties of the system under scrutiny. Fundamental limits to the controllability of an open system can be formalized in terms of the law of requisite variety, which is derived from the second law of thermodynamics and suggests that establishing correlations between the system under scrutiny and a controller is beneficial. These concepts are briefly reviewed, and the chances, challenges and potential gains arising from the generalisation of such a framework to the quantum domain are discussed.  In particular, recent findings in quantum information theory unveiled a new kind of quantum correlations called quantum discord. We conjecture a quantitative link between quantum correlations and controllability, i.e. quantum discord may be employed as a resource for controlling a physical system.    
\end{abstract}

\maketitle

\section{What is cybernetics?}

\subsection{The problem}

The word {\it cybernetics} ({\it ``The art of steersmanship''}) was coined by Norbert Wiener in 1948 \cite{wiener} to define a cross-disciplinary research field aimed at studying regulatory phenomena in a broad range of contexts, from engineering to biology, from finance to cognitive and social sciences. {\it ``The art of steersmanship [...] stands to the real machine -- electronic, mechanical, neural, or economic -- much as geometry stands to real object in our terrestrial space; offers a method for the scientific treatment of the system in which complexity is outstanding and too important to be ignored.''}\cite{ashby} This statement by W. Ross Ashby highlights the trans-disciplinary vocation of cybernetics, a meta-theory for describing common features of complex systems.

One may wonder what exactly we mean by ``complex''. This is a term which enjoys many possible interpretations. We refer the reader to a reference textbook \cite{politi} and an  extended analysis of the state-of-the-art in quantifying complexity \cite{lena}.  To establish a solid ground for the discussion, let us state that a complex system is an aggregate of many parts which interact nontrivially with each other. By nontrivial interaction we refer to correlations which allow the global system to  behave in a qualitatively different way with respect to the parts considered separately. In Aristotle's poetic words, complexity emerges when {\it ``The whole is other than the sum of its parts''}. The actual partition of the system is usually determined by the problem particulars, for example the spatial separation between the components of the system, or the role they play in a specific information processing protocol. Cybernetics focuses mostly with the latter case.

\begin{figure}[h!]
\centering
\includegraphics[scale=.7]{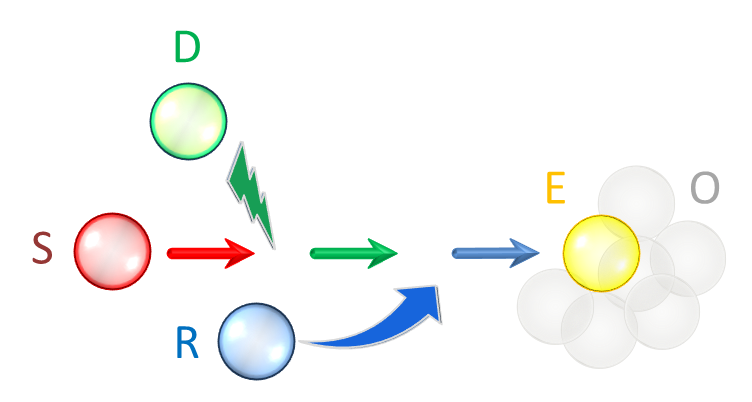}
\caption{(Colors online) Scheme of a regulatory process. An experimentalist aims to drive the system $S$ (red ball) into a desirable target configuration $E$ (yellow) out of all possible outcomes $O$ (light gray). The environmental noise is modelled by means of a second system $D$ (green) which disturbs $S$. A third system $R$ (blue) is available in the laboratory to correct the evolution of $SD$ in order to drive $S$ to the target state $E$.  \label{fig1}
}
\end{figure}

The general problem we are investigating can be formalized as depicted in Fig.\ref{fig1}. A system is initially prepared in a state $S$. An observer, or the system itself, wants to regulate the dynamics of the system in order to reach their  expected outcome, or goal, $E$.  This entails to balance or counteract the typically detrimental action of an external disturbance (e.g.~the environment or an adversarial agent) $D$ by applying a control, or regulation strategy, $R$ (either by accessing an ancillary system, or by internal mechanisms). The aim is to (self-)drive the system into a desirable final state $E$, e.g. to send a living system into a state in which it is still alive and healthy. 
  In the context of cybernetics, the specific nature of the information content and the physical properties of the information carriers do not have any relevance. The roles of system, regulator and disturbance are dictated by the experimentalist or by constraints inherent to the problem of interest.

\subsection{Quantum Cybernetics vs Quantum Control}

One may notice that Fig.\ref{fig1} is nothing but a protocol where information is processed  in order to drive the system into a target state, i.e.~ towards an objective. Such a problem has been widely discussed in the control theory, which has been proven highly successful in the last decades, both in the classical and in the quantum regimes \cite{survey,revrabitz}. In particular,  several feedback control strategies, where information is obtained by a measurement or a coherent interaction and then employed by the controller to implemement the appropriate driving dynamics to the system $S$, have been proposed and succesfully applied \cite{wiseman,mabuchi,gough,james,viola,lloyd}. However, they definitely departed from the cybernetics approach, which tackles the problem from an information theory viewpoint. The question is to set general prescriptions to  determine the minimal requirements for a successful regulation  and to explore the limits imposed by the fundamental law of physics, in particular the second law of thermodynamics. In fact, an information-theoretic analysis of classical state regulation has been provided \cite{touchette,touchette2,engell}, but a full treatment in the quantum mechanical scenario is missing. In a parlance which is familiar to information theorists, we may then ask what is the {\it resource} for quantum regulation. The peculiar ability of quantum systems to store and process information harnessing nonclassical features such as coherence and quantum correlations (not limited to entanglement), suggests that a quantum controller may be intrinsically more efficient than a classical one in the regulation of open, quantum or classical complex systems. In particular, we are interested in cooperative effects between the regulator and the disturbance, i.e. when their interaction makes the difference between a hostile and a helpful environment.  
We note that the protocol in Fig. \ref{fig1} resembles information processing tasks as quantum teleportation \cite{telep}, remote state preparation \cite{remote}, and quantum state merging \cite{merging}. All these protocols can be rethought as state driving problems under different constraints, and it is well known that quantum correlations play a decisive role in their optimal realisation \cite{horo,dakic,mergint,hyb}. Therefore it seems sound and interesting to us to investigate if a more general statement about the role of quantum correlations in quantum control is possible.  The controllability of the system may be benchmarked by how much correlations need to be created, i.e. what is the optimal trade off  between creation of correlations when information goes from the system to the controller and correlation consumption during the feedback step.   Also, recent results in quantum information theory led to refine the law of thermodynamics for individual quantum systems \cite{oscar,opp}, thus calling for shaping the limits to controllability in such a scenario. To do that, a quantum cybernetics, i.e. an information-theoretic study of quantum state driving,  is required.


The paper is organized as follows. We will discuss the classical limitations to successful regulation in Section \ref{classical}, which are summarized by the surprisingly simple law of requisite variety, originally introduced by Ashby \cite{ashby} and then rediscovered and extended in more recent years \cite{shalizi,touchette,touchette2,engell}. In Section~\ref{quantum}, we comment on potential quantization strategies for the regulation protocol of Fig.~\ref{fig1}. It is then legit to ask if Nature provides us with examples of optimal (quantum) regulation, e.g. in the {\it par excellence}  complex systems, i.e. the biological ones. We discuss the exploitability of an information theory of classical and quantum control to self-regulating and biological systems in Section~\ref{bio}. We draw our conclusions in Section~\ref{fine}.

\section{Regulation of an open system}\label{classical}
\begin{figure}[h!]
\centering
\includegraphics[width=\linewidth]{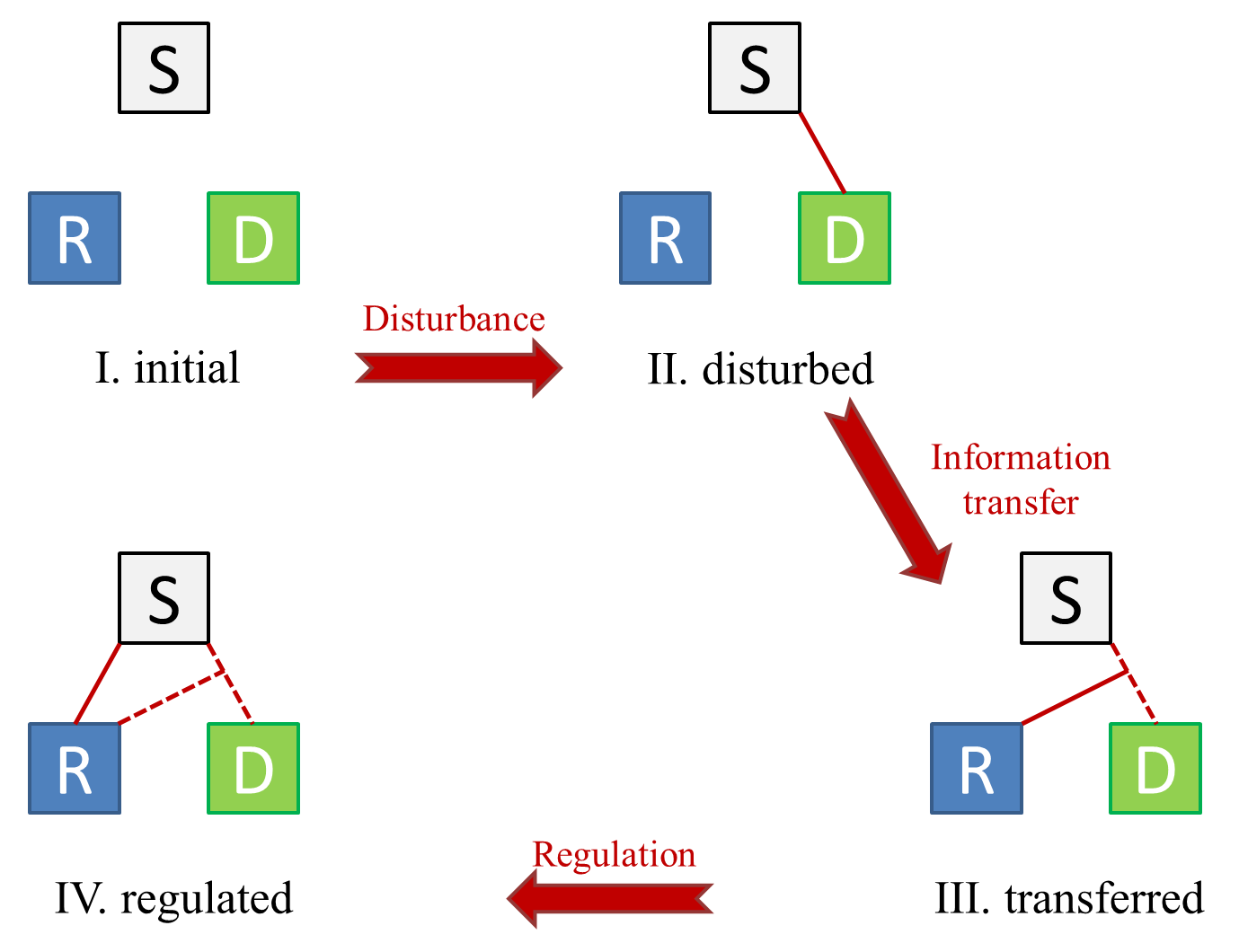}
\caption{(Colors online) Tripartite composite system, consisting of the principal system $S$, an environment $D$ and regulator $R$. The dynamics are governed by the interaction between the subsystems. \label{fig2}
}
\end{figure}

In general, we consider a tripartite composite system, consisting of the principal system $S$, an environment $D$, whose action into the system provides the disturbance, and a regulator $R$, whose interaction with the system (and possibly with the environment) provides the regulation, see Fig.~\ref{fig2}. As we are only interested in the system $S$ and the interaction with the other two components, the setting is complex in the sense that the relevant dynamics is largely determined by the correlations between the subsystems. In particular the combined action of both disturbance and regulation can drive the system to goals which neither of the respective bipartite interactions are able to achieve on their own, as has been shown, e.g. in \cite{RS,RS2}. However, a complexity measure based on the number of reachable states (with the focus on a possible increase due to the tripartite interaction) misses out a crucial point of control setting: Not the bare number of reachable states are relevant, but whether the desirable states are reachable. It then turns out that  Fig.~\ref{fig1} is slightly misleading, as it suggests a time-ordering and separating of disturbance and regulation as well as implying basically an error correction mechanism. In the most general case, in fact, for realistic open quantum systems,  it is reasonable to assume instead that disturbance and regulation interact in parallel with the system. Or even, when e.g. decoherence-free subspaces are employed for control tasks \cite{lidar}, that the regulation acts before (i.e. quicker than) the disturbance. (Quantum) cybernetics is also not restricted to control settings, where the task for the regulator consists solely of inverting the disturbance. Nevertheless, for the sake of clarity, we stick in the figures to this exemplary setting.

\subsection{The law of requisite variety}
The state of a (classical) system can be described by a set of values of its relevant variables. Given a set $S$, the logarithm (in base $2$) of the number of distinguishable elements in the set defines the {\it variety} ${\cal V}_S$ of the set. For instance, the set $\{a,a,b,c,a,c,b,b\}$ has variety $\log_2 3$. Having in mind the prototype of Fig.~\ref{fig1}, if we have a system $S$ in a certain initial state, a disturbance induces a set $D=\{d_1,d_2,\ldots\}$ of possible undesired actions. The regulatory mechanism, in turn, is able to produce a set $R=\{r_1,r_2,\ldots\}$ of responses. The final state of the system is therefore determined by a payoff matrix of possible outcomes $O=(o_{ij})$, corresponding to each pair $d_i, r_j$ (this was introduced in \cite{games} for studying games and economic strategies). For example, if we are driving our car, all kind of disturbances can happen. If $d_1$ is: ``a person suddenly crosses in front of us'', then a response $r_1$: ``do nothing'' leads to $o_{11}$: ``the person is ran over'', while a better regulation $r_2$: ``brake'' leads to $o_{12}$: ``safety''. Similarly, if $d_2$ is ``it starts raining'' the best regulatory action is to switch on the wipers, which leads to ``safety'' once more; and so on. Eventually, the matrix $O$ of possible outcomes can be quite big if one considers all possibilities for $D$ and $R$. In typical complex phenomena, the desired, or expected outcome $E$ is only a small subset of all that can happen. In our car we just want to drive safely and reach our destination. If we define $E = \{$``safety''$\} \subset O$, then we will be able to achieve our goal provided that, for every $d_i$, there exists at least one action $r_j$ which leads to safety.  Therefore, the role of the regulator is to reduce the achievable variety in the outcomes $O$. It is intuitive to see that, in order to do so, the regulator itself has to have a sufficiently high variety. That is, we need enough controls in our car to counteract the various mishaps which might occur.
The law of requisite variety formalizes this quantitatively in a simple inequality \cite{ashby}:
\begin{equation}
\label{req}
{\cal V}_O \geq {\cal V}_{D}-{\cal V}_R\,.
\end{equation}
Thus,  the entropy of the regulator must be bigger than a
function of the entropies of the disturbing system and the potential final states of the controlled system. A useless controller, which always responds with the same regulatory action (${\cal V}_R=0$) will result in an outcome with at least as much variety as the one of the disturbance $D$. On the other hand, a perfect controller is able to release a counteraction for every disturbance (${\cal V}_R={\cal V}_D$) so that, ideally, the possible outcomes are reduced to the expected goal, $O=E$, with minimal variety.

As it stands, the law of requisite variety is formulated at a very general, abstract level. What is the connection with information theory? It comes naturally once we make a further assumption, that the process under scrutiny can be repeated $n$ times, $n\rightarrow \infty$. This has been the traditional setting for communication theory. In this context, we can think of $D$, $R$, and $O$ as three random variables. The variety of a statistical variable $X$, which can assume outcomes $\{x_i\}$ with probabilities $\{p_i\}$, can then be interpreted as its {\it entropy}
\begin{equation}
\label{shan}
{\cal H}(X) = -\sum_i p_i \log_2 p_i
\end{equation}
as adopted by Shannon \cite{shannon}, which laid the mathematical foundation for information theory in a probabilistic framework. In such asymptotic scenario, if a message to be transmitted consists of $n$ independent and identically distributed (i.i.d.) random variables $X^{\otimes n}$, then such a message can be noiselessly  encoded in a string of bits of length at least $n {\cal H}(X)$. The bit, an entity which can take two values ($0$ or $1$), physically represented e.g.~by a coin or a light switch,  is the fundamental unit of information. On this hand, Ashby's law appears as a generalization of
the Shannon equivocation theorem (Th. 10 of \cite{shannon}): In a typical communication setup, a sender wants to transmit a message to a
receiver, but the message is sent down a noisy channel. A
regulator would then be another channel used as a corrective
tool to filter out the undesired randomness from
the received message.

Using entropy to revisit the law in Eq.~(\ref{req}) for three random variables $D,R,O$, we recall as previously noticed that, if the regulator has a fixed realization, then all variety (entropy) of $D$ is retrieved at $O$. One can write, using conditional entropies, ${\cal H}(O|R) = {\cal H}(D|R)$. Exploiting the properties of the entropy one can easily show that \cite{ashby}
\begin{equation}\label{req2}
{\cal H}(O) \geq {\cal H}(D)-{\cal H}(R)+{\cal H}(R|D)
\end{equation}
which is a restatement of the law of requisite variety (upon observing that ${\cal H}(R|D)=0$ if the regulator has a deterministic action in response to a disturbance). This has been independently rederived  in the contexts of control theory \cite{touchette,touchette2,engell} and computational mechanics \cite{shalizi}.
In Eq.~\eqref{req2} the term ${\cal H}(R)-{\cal H}(R|D)$ is of course the mutual information between regulation and disturbance, i.e. the information the regulator can make use of, in order to drive the system to the desired goal. In fact, as shown in \cite{Slaw}, Ashby's law of requisite variety is essentially a formulation of the second law of thermodynamics. Nevertheless, it focuses on an important aspect, as it tells us how much information the regulator must be able to store for a successful regulation. In terms of resources, this is also a question of complexity, or more specifically of correlations between the three parties of the protocol.
For a classical system, as stated, the conditional entropy ${\cal H}(R|D)$ is in the best case zero. This, however, does not hold in the quantum case. It is then natural to wonder: Can quantum correlations or other signatures of quantumness be exploited to 
improve the performance of the regulator?



\section{Perspectives for requisite variety in the quantum domain}\label{quantum}

We now discuss possible extensions to the quantum case of the regulation protocol.  These would set general limitations on the controllability
of a quantum system. Inspired by the successes of quantum control and quantum information processing, one would expect to find that quantum regulators are more efficient than classical ones. The main issue to tackle is how to define variety in the quantum domain.

\subsection{Von Neumann entropies}

The most obvious way to define variety is by replacing the Shannon entropy ${\cal H}$ with the von Neumann entropy ${\cal S}$ and the random variables replaced by the quantum states of the systems involved in the regulatory process.
Although apparently innocent, such a step already features nontrivial subtleties. First of all, the notion of conditional entropy has to be carefully defined. We have two possibilities for ${\cal H}(R|D)$ in Eq.~(\ref{req2}). One is to use formally the same expression as in the classical case, ${\cal S}(R|D) = {\cal S}(R,D) - {\cal S}(D)$. However, this quantity can be negative. This can happen in particular when the quantized systems $D$ and $R$ are in an entangled state. The negativity of the conditional quantum entropy has been operationally interpreted in the context of quantum state merging \cite{merging}, an important primitive of quantum Shannon theory \cite{wilde}, and in quantum thermodynamics \cite{oscar}. The other possibility is to imagine that, in order to learn about the occurred disturbance, $R$ performs a measurement $\{\Pi_j^D\}$ on $D$. The role of the observer in the regulation, precisely the disturbance induced by applying a general quantum map to a given state,  can be neglected in a classical scenario, but it plays a decisive role in the quantum case.

In this picture, the conditional entropy, optimized over all possible measurements (to single out the least disturbing one), can be written as
\begin{equation}
\tilde{{\cal S}}(R|D) = \inf_{\{\Pi_j^D\}} {\cal S}(R|\{\Pi_j^D\}).
\end{equation}
The latter quantity  is always nonnegative. It turns out that the two quantities ${\cal S}(R|D)$ and $\tilde{{\cal S}}(R|D)$ coincide if and only if $D$ is effectively classical; when this does not happen, the regulator and the disturbance display quantum correlations as revealed by the so-called {\it quantum discord} \cite{zurek,genio}. Quantum discord and in general quantum correlations between $R$ and $D$ are one strong element which can mark a departure from the classical paradigm studied by Ashby. In particular, if $D$ and $R$ share quantum correlations we can expect by looking at Eq.~(\ref{req2})---rewritten with von Neumann entropies---that such correlations constitute a further, genuinely quantum resource to lessen the variety in the outcome $O$, in addition to the purity of the regulator. This is especially true when $R$ and $D$ are entangled, as ${\cal S}(R|D)$ can be negative as observed before.

\subsection{Smooth Renyi entropies}

Shannon and von Neumann entropies are meaningful figures of merit for asymptotic information theory, i.e., for ergodic systems. However,
Ashby himself recognized that ergodicity is too often not a realistic feature of cooperative living or social phenomena \cite{ashby}. Much more recently, theorists in classical and quantum information theory have also (independently) recognized that the paradigm of i.i.d. messages is not necessarily respondent to common practice. It is rather more natural to consider {\it one-shot} scenarios: the sender encodes a message in a single system, transmits it down a single channel, possibly resorts to another single additional regulative channel, and the receiver receives and decodes the message in a single run. If more trials of a process are repeated, it is equally unrealistic to assume that they are completely independent. In data transmission, as well as in biological processes, the different runs are typically correlated. In such a more general situation, the association between variety and the Shannon/von Neumann entropies is not correct anymore, and we need to resort to the primitive formulation of Ashby's law as given by Eq.~(\ref{req}).

Fortunately, a more general framework to define and quantify information in the non-i.i.d. setting, in both classical and quantum scenarios, is available and makes use of so-called smooth Renyi entropies \cite{smooth}. Therefore another, perhaps more informative, avenue towards quantum cybernetics suitable for quantum and nanoscale systems goes through an alternative formulation of variety in terms of such entropies, in particular the smooth min- and max-entropies, which have a nice operational intepretation in one-shot information theory and cryptography \cite{smoothinterpret,oneshot}. Let us just recall that Renyi entropies are a family of additive entropies defined as
${\cal S}_\alpha(\rho) = (1-\alpha)^{-1} {\rm{Tr}}[\rho^\alpha]$. Min-entropy corresponds to $\alpha\rightarrow\infty$ and max-entropy to $\alpha \rightarrow 1/2$. Very recently, adopting the formalism of smooth Renyi entropies, it has been shown that there exists a set of many second laws in quantum thermodynamics \cite{opp}. We can then expect to recover equivalently many laws of requisite variety in the quantum domain. Let us note that measures of quantum correlations can be defined in the case of smooth Renyi entropies as well \cite{smoothdisc}, which gives us in principle all the necessary tools to quantify quantum advantages and the role of nonclassical effects for enhanced regulation.

 \subsection{Cost functions}

Finally, a remark is in order. The law of requisite variety is essentially an inequality between probability distributions. One may wonder if this minimal description is sufficient to describe efficiently any regulation protocol. In general, even if the goal is to achieve a pure state of the system (and this is not always true), the focus on entropy reduction both in the ergodic and non-ergodic case does not distinguish between different pure states.  Therefore, while focusing on entropy reduction seems the standard approach to employ information theory to control problems \cite{touchette,touchette2,jacobs},  this is arguably not universally sufficient for control even in the classical case \cite{engell,belavkin}. Optimal regulation should then be benchmarked by the minimisation of an appropriate, experimentally appealing cost function. For instance, for practical purposes, a fidelity measure appropriate for the regulative task under investigation appears to be a reasonable resort \cite{doherty}. 
This will be the subject of further investigation \cite{natureglory}.

\section{Classical and quantum  biology in the cybernetics paradigm}\label{bio} 
The regulation of biological systems and their interplay with the environment appear as the ideal testbeds for the law of requisite variety.
Plenty of examples can easily be concocted which fit
into this paradigm. In any living system, the desired outcome
is to stay alive and healthy, the disturbance can be
caused by toxins such as bacteria and viruses, and suitable
anti-toxins act as the regulators.  The question we raise is if the law
of requisite variety, or one of its declinations, may serve
as a general design principle for biological complexes.
Here we consider two specific classes of problems which may be interpreted as examples of regulation protocols.

The first case study is related to chemotaxis, i.e. the dynamics of microscopic bacteria based on information about gradients of the concentration of specific chemical elements in the environment (e.g. searching for food). Search algorithm inspired by chemotaxis for macroscopic devices (robots) working with incomplete information about the environment (infotaxis) have been developed \cite{infotaxis}.  At the same time, an information theoretic analysis of chemotaxis as a self-regulation protocol has been recently proposed \cite{auletta}.  It should be then of great interest to determine if bacteria perform at the limits imposed by Ashby's law, as well as to assess the performance of bio-inspired devices against the information-theoretic limits to controllability. If an optimal quantum controller can overcome such limits, then quantum mechanics may find a surprising new functional role in robotics.

 The other intriguing question is if biological systems exploit non-trivial
quantum effects for their optimal regulation and adaptation.
Three main biological processes are currently under investigation by quantum physicists: the energy transport mechanisms regulating photosynthesis \cite{fmo}, the magneto-reception system of birds \cite{birds}
and the olfactory sense \cite{turin}.  Focusing in particular on the first two examples, recent experimental evidence and theoretical modelling suggest that coherence (in the case of light-harvesting organisms such as the Fenna-Matthew-Olson complex \cite{fmo}) and entanglement (in the case of the radical pair model for the avian compass e.g.~in the European robins \cite{birds}) are exploited by living systems to optimize their biological processes in the presence of a decohering environment.
These case studies have triggered in the last decade the dawn of quantum biology as a multidisciplinary research field \cite{qbio}.

Seen in the light of what presented in the previous section, these are
 clear examples of regulatory phenomena: organisms pursuing a physiological function, subject to
external disturbance $D$, and responding with a (self)-regulatory action $R$ so that their outcome $O$ is kept
at sufficiently low variety, thus ensuring that the expected goal $E$ (e.g. transporting a photon from the photoreceptor to the reaction centre in the case of photosynthetic complexes, or maintaining an accurate navigation route for migratory purposes in the case of birds) is achieved with the highest possible
chance. What is remarkable, is that we are dealing with undoubtedly complex macroscopic systems
which would traditionally be ascribed to the classical domain, and are certainly in contact with classical environments. Yet,
in a multibillion-year stint of evolution, they appear to have developed effective quantum strategies for
their optimal regulation.

A current challenge is to understand the key principle(s) underpinning such possible quantum effects in biology. In particular, it is pivotal to identify the resource which enables biological systems to control and exploit the interplay
with the environment. It is known that biological processes are optimized by intermediate levels of coherence, i.e. too much coherence can be detrimental. Thus, we should search for a more-elusive-than-coherence resource. The
answer may be in the structure, i.e. the degree of organisation of the system itself, which allows the complex to self-regulate its dynamics.

\section{Conclusions}\label{fine}
Control theory has a long and glorious history of successes \cite{survey,revrabitz}, yet we have still incomplete knowledge about a general design principle of optimal controllability in open quantum systems.  Combining the quantum control rationale and the latest results in quantum information   may lead to the establish the ultimate, general, quantitative limits to the controllability of quantum systems.  An important finding of classical cybernetics is the law of requisite variety:  the purity of the controller and the degree of correlations it can establish with the system determines the controllability of the latter, independently of  its peculiar chemico-physical
properties. Quantum cybernetics (cf. \cite{belavkin}) will provide the framework for a fundamental study of the role that quantum effects and quantum correlations play in the regulation of open quantum and classical systems. Furthermore, it will enable a rigorous treatment of self-regulating quantum systems, where regulator and environment are effectively the same physical object, whose interaction with the principal system has different effects for different timescales. As the framework in Fig.\ref{fig1} is independent of the particulars of the considered problem, it is applicable to a number of apparently unrelated phenomena as bacteria infotaxis and photosynthesis.

The current technological advances in the manipulation of single quantum systems demand for investigations on the ultimate limits to the controllability of physical systems imposed by quantum mechanics.  When reading the original works of Ashby and Wiener from half a century ago, one can be delightfully surprised by  the modern flavour of their insights: their line of thinking resonates very closely with the state-of-the-art research and challenges in contemporary quantum information theory and technology.

With this article, we hope to have stimulated the interest of the reader in the topic and to have proved it worthwhile of attention.  Once quantum cybernetics will be fully developed, experimental proof-of-concepts of the ensuing limitations on regulative processes may be in reach of current technological possibilities. Also, firm answers to at least some of the following questions will be provided: How much I have to correlate the controller to the system under investigation to obtain a certain degree 
of controllability? Is the quantum treatment of the problem significatively different from the classical one? Does 
quantum discord, a recently discovered and very debated quantum feature, help controlling a quantum system? Do 
complex systems in Nature exploit this supposed quantum advantage? Controllability is a task regulated by the law 
of thermodynamics. How "one shot" quantum control works?
  
      Discovering the ultimate strategies to manipulate single quanta may translate into the ability to control other kinds of systems which are far removed from the nanoscale regime, i.e. certain social systems \cite{social}, as an audacious mind precognized almost a century ago \cite{majo}.

\section*{Acknowledgements}
We acknowledge fruitful discussion with F.~Carusela, L.~A.~Correa, S.~De Martino, S.~Lee, P.~Liuzzo Scorpo, and S.~Lloyd. This work was supported by the Foundational Questions Institute (Grant No.~FQXi-RFP3-1317), the ERC StG GQCOP (Grant No.~637352), the UK Engineering and Physical Sciences Research Council (Grant No.~EP/L01405X/1) and the Wolfson College, University of Oxford and The University of Nottingham Staff Travel Prize.

\end{document}